\documentclass[11pt]{article}
\usepackage{amsfonts}
\usepackage{graphicx}
\usepackage{epstopdf}
\usepackage{epsfig}
\usepackage{amssymb}
\usepackage{setspace}
\usepackage{caption}
\usepackage{amsfonts}
\usepackage{color}
\usepackage{amsmath}
\usepackage{float}
\usepackage{comment}
\usepackage{subcaption}
\newcommand {\be}{\begin{equation}}
\newcommand {\ee}{\end{equation}}
\newcommand {\bea}{\begin{array}}
	
	\newcommand {\eea}{\end{array}}

\textwidth=6.5in \hoffset=-.75in \voffset=-1in \numberwithin{equation}{section}
\numberwithin{figure}{section}

\begin{document}

\begin{titlepage}
	\vspace{1cm}
	\begin{center}
		{\Large \bf {Magnetized black holes in Kaluza-Klein theory and the Kerr/CFT correspondence}}\\
	\end{center}
	\vspace{2cm}
	\begin{center}
		\renewcommand{\thefootnote}{\fnsymbol{footnote}}
		Haryanto M. Siahaan{\footnote{haryanto.siahaan@unpar.ac.id}}\\ Program Studi Fisika, Universitas Katolik Parahyangan,\\
	Jalan Ciumbuleuit 94, Bandung 40141, Indonesia
		\renewcommand{\thefootnote}{\arabic{footnote}}
	\end{center}
	
	\begin{abstract}
		
In this work, we examine the Kerr/CFT correspondence for magnetized black holes arising from Kaluza--Klein theory, demonstrating that Kerr/CFT holography persists beyond the traditional Einstein--Maxwell framework. Notably, unlike in the Einstein--Maxwell case, the massless neutral scalar field equation here is fully separable into radial and angular parts. This separability reveals a hidden conformal symmetry in the near--horizon, low--frequency regime, providing further support for the robustness of Kerr/CFT dualities in extended gravitational theories.
		
	\end{abstract}
\end{titlepage}\onecolumn
\bigskip

\section{Introduction}
\label{sec.intro}

Magnetized black holes have long been a central topic in gravitational physics, owing to the ubiquity of strong magnetic fields in astrophysical environments around the horizon \cite{Kumar:2014upa}. Foundational studies by Wald \cite{Wald:1974np} and Ernst \cite{Ernst:1976mzr}, later extended by Wild and Kerns \cite{Wild:1980zz} and Aliev and Galtsov \cite{Aliev:1989wz}, showed that an external uniform magnetic field can significantly modify electromagnetic interactions near a black hole, deform the horizon geometry, and impact its thermodynamic behavior. More recently, horizon-scale observations by the Event Horizon Telescope collaboration \cite{EventHorizonTelescope:2021bee,EventHorizonTelescope:2021srq} have confirmed that strong magnetic fields permeate the plasma surrounding supermassive black holes, playing a key role in fueling accretion processes and launching relativistic jets. These developments highlight the continued importance of studying magnetized black holes to deepen our understanding of high-energy astrophysical phenomena \cite{Narayan:2021qfw,Hou:2022eev,EventHorizonTelescope:2021btj,Junior:2021dyw}.

Holographic approaches have opened a new frontier in black hole physics by enabling the study of gravitational systems through their dual field theory descriptions. A particularly prominent example is the Kerr/CFT correspondence \cite{Guica:2008mu}, which relates the near-horizon geometry of extremal rotating black holes to two-dimensional conformal field theories (CFTs). This duality provides a microscopic framework for understanding macroscopic black hole properties—such as entropy and thermodynamics—thereby deepening our insight into quantum gravity and holography \cite{Compere:2012jk,Hartman:2008pb}. A key element of this correspondence is the enhanced symmetry near the horizon of extremal rotating black holes, first identified by Bardeen and Horowitz \cite{Bardeen:1999px}. The Kerr/CFT framework exploits this symmetry by constructing a Virasoro algebra via the Brown--Henneaux method \cite{Brown:1986nw}, which in turn allows the Cardy formula to accurately reproduce the Bekenstein--Hawking entropy. This powerful holographic principle has since been successfully extended to a wide class of rotating and charged black holes in diverse gravitational theories \cite{Chen:2010as,Chen:2010jj,Sakti:2019zix,Sakti:2020jpo,Ghezelbash:2009gf}, greatly expanding its reach and relevance.

A particular extension of the Kerr/CFT correspondence arises within the framework of Kaluza--Klein (KK) theory, which aims to unify gravity and electromagnetism through the compactification of an extra spatial dimension \cite{Klein:1926tv}. KK black holes feature both gauge fields and scalar dilaton fields, resulting in richer gravitational structures that respond dynamically to external electromagnetic influences \cite{Breitenlohner:1987dg}. Similar to the case in Einstein--Maxwell (EM) theory, magnetized black hole solutions can also be constructed in the Kaluza--Klein setting. Yazadjiev \cite{Yazadjiev:2013qna,Yazadjiev:2013hxa} demonstrated how such magnetized KK black holes can be obtained and analyzed their fundamental properties, including the computation of conserved quantities and the formulation of corresponding thermodynamic relations.

In recent years, there has been sustained interest in extending the Kerr/CFT correspondence beyond the EM paradigm \cite{Siahaan:2024ilq,Sakti:2023fcu,Sakti:2022txd}. The successful application of Kerr/CFT holography to extremal magnetized Kerr black holes in EM theory \cite{Siahaan:2015xia,Astorino:2015naa} naturally motivates the question of whether this duality continues to hold in more generalized frameworks, such as KK theory. Notably, the magnetized Kerr/CFT correspondence has been extended to a wider class of EM black holes incorporating the NUT parameter \cite{Siahaan:2021ags,Siahaan:2021uqo,Ghezelbash:2021xvc}. In those studies, analyses of the near-horizon geometry demonstrated that the holographic correspondence remains valid even in the presence of strong magnetic fields. Explicit computations of the central charge, associated with the Virasoro algebra arising from diffeomorphisms near the horizon, revealed consistency between the macroscopic Bekenstein--Hawking entropy and the microscopic entropy derived via the Cardy formula. Interestingly, some of these results exhibited negative central charges, suggesting that the corresponding dual CFT may be non-unitary. This raises an important theoretical question: does a similar non-unitary feature, indicated by a negative central charge, also emerge in the magnetized Kerr black holes of KK theory?

This paper aims to address the questions raised above by extending the Kerr/CFT correspondence to magnetized Kerr black holes in KK theory. Our first objective is to examine the conformal symmetry associated with the near-horizon geometry of an extremal magnetized Kerr black hole in KK theory. Identifying this symmetry enables the application of the Kerr/CFT holographic framework, through which we compute the corresponding microscopic entropy. Beyond the extremal case, we also investigate the presence of hidden conformal symmetry by analyzing the radial part of the Klein--Gordon (KG) equation for a test scalar field near the black hole. Notably, in EM theory, the KG equation fails to separate under general magnetic field configurations, thereby limiting the construction of hidden conformal symmetry. In contrast, we explore whether the magnetization in KK framework allows for full separability, thus extending the reach of holographic methods to non-extremal magnetized black holes.

The paper is organized as follows. In the next section, we provide a short review of the construction and key features of magnetized black hole solutions in KK theory. Section~\ref{sec.thermo} is devoted to a discussion of the black hole's thermodynamic properties. In section~\ref{sec.nearhor}, we establish the Kerr/CFT holography for the extremal magnetized Kerr black hole within KK theory. Subsequently, section~\ref{sec.hiddenconf} extends this analysis to the non-extremal case, emphasizing the hidden conformal symmetry associated with the separability of the KG equation. Finally, we summarize our findings in the conclusion section. Throughout this paper, we adopt natural units with \( c = G_N = \hbar = 1 \).

\section{Review on the magnetization of Kaluza-Klein black holes}

In the following, we employ three sets of fields to describe different configurations: the vacuum seed solution, the KK configuration, and its magnetized extension. These are denoted by \(\left\{ \tilde{g}_{\mu \nu} \right\}\), \(\left\{ \bar{g}_{\mu \nu}, \bar{A}_\mu, \bar{\Phi} \right\}\), and \(\left\{ g_{\mu \nu}, A_\mu, \Phi \right\}\), respectively. The sets \(\left\{ \bar{g}_{\mu \nu}, \bar{A}_\mu, \bar{\Phi} \right\}\) and \(\left\{ g_{\mu \nu}, A_\mu, \Phi \right\}\) obey the effective action of four-dimensional KK theory \cite{Aliev:2008wv}, given by
\be \label{eq.action}
S = \int d^4x\,\sqrt{-g}\,\Bigl[ R - 2\,\partial_\mu \Phi\,\partial^\mu \Phi - e^{2\sqrt{3}\,\Phi}\,F_{\mu\nu}F^{\mu\nu} \Bigr],
\ee 
where \(R\) is the four-dimensional Ricci scalar, \(\Phi\) is the dilaton field, and \(F_{\mu\nu} = \partial_\mu A_\nu - \partial_\nu A_\mu\) is the field strength of the \(U(1)\) gauge field \(A_\mu\). Varying the action with respect to the metric, dilaton, and gauge field yields the corresponding equations of motion.
\be\label{eq.Rmn} 
R_{\mu\nu}=2\,\partial_\mu\Phi\,\partial_\nu\Phi+2\,e^{2\sqrt{3}\,\Phi}\Bigl[F_{\mu\alpha}F_\nu^{\ \alpha}-\tfrac{1}{4}\,g_{\mu\nu}\,F_{\alpha\beta}F^{\alpha\beta}\Bigr]\,,
\ee
\be\label{eq.Am}
\nabla_\mu\Bigl(e^{2\sqrt{3}\,\Phi}\,F^{\mu\nu}\Bigr)=0\,,
\ee
\be \label{eq.Phi}
2\,\nabla^2\Phi-\sqrt{3}\,e^{2\sqrt{3}\,\Phi}\,F_{\alpha\beta}F^{\alpha\beta}=0\,.
\ee

A well-established method for generating solutions to the KK field equations involves starting from a known vacuum solution of four-dimensional Einstein gravity as a seed. In this procedure, one first uplifts the four-dimensional vacuum metric \(\tilde{g}_{\mu\nu}\) to five dimensions via the ansatz
\be \label{eq.ds5}
ds_5^2 = \tilde{g}_{\mu\nu}\,dx^\mu\,dx^\nu + dz^2 = G_{MN}\,dx^M\,dx^N,
\ee 
where the four-dimensional coordinates are \(x^\mu = \{t, r, \theta, \phi\}\), and the extended five-dimensional coordinates are \(x^M = \{x^\mu, z\}\). The resulting five-dimensional metric \(G_{MN}\) satisfies the vacuum Einstein equations \(R_{MN} = 0\), which in turn implies that the original four-dimensional seed metric \(\tilde{g}_{\mu\nu}\) is Ricci-flat, i.e., it satisfies \(R_{\mu\nu} = 0\).

To obtain a new solution within KK theory, one performs a Lorentz boost in the fifth dimension \cite{Aliev:2008wv}, defined by the coordinate transformation
\be 
\begin{aligned}\label{eq.boost}
	dt &\to \cosh\alpha\, dt + \sinh\alpha\, dz\,,\\[1mm]
	dz &\to \cosh\alpha\, dz + \sinh\alpha\, dt\,,
\end{aligned}
\ee
where \(\alpha\) is the boost parameter. Applying this transformation to the five-dimensional metric results in the line element
\be 
ds_5^2 = H^{-1}\,\bar{g}_{\mu\nu}\,dx^\mu dx^\nu + H^2 \left(dz + 2A_\mu\,dx^\mu\right)^2\,,
\ee
where the function \(H\) is defined as \(H^2 = \cosh^2\alpha + \tilde{g}_{tt}\,\sinh^2\alpha\). The resulting four-dimensional metric \(\bar{g}_{\mu\nu}\) can be written in the tetrad formalism as
\be 
ds^2 = \eta_{(i)(j)}\,\boldsymbol{\omega}^{(i)}\,\boldsymbol{\omega}^{(j)}\,,
\ee
with the one-forms given by
\[
\begin{aligned}
\boldsymbol{\omega}^{(0)} &= \sqrt{\frac{\tilde{g}_{tt}}{H}}\left(dt + c\,\frac{\tilde{g}_{t\phi}}{\tilde{g}_{tt}}\,d\phi\right),\\[1mm]
\boldsymbol{\omega}^{(1)} &= \sqrt{H\,\tilde{g}_{rr}}\,dr\,,\\[1mm]
\boldsymbol{\omega}^{(2)} &= \sqrt{H\,\tilde{g}_{xx}}\,dx\,,\\[1mm]
\boldsymbol{\omega}^{(3)} &= \sqrt{\frac{H}{\tilde{g}_{tt}}\left(\tilde{g}_{\phi\phi} \tilde{g}_{tt} - \tilde{g}_{t\phi}^2\right)}\,d\phi\,,
\end{aligned}
\]
and the Minkowski signature is encoded in \(\eta_{(0)(0)} = -1\), \(\eta_{(1)(1)} = \eta_{(2)(2)} = \eta_{(3)(3)} = 1\). The corresponding non-gravitational fields that, along with \(\bar{g}_{\mu\nu}\), satisfy the equations of motion (\ref{eq.Rmn})--(\ref{eq.Phi}) are as follows. The gauge field takes the form
\be 
A_\mu\,dx^\mu = \frac{\sinh\alpha\left[\left(1 + \tilde{g}_{tt}\right)\cosh\alpha\,dt + \tilde{g}_{t\phi}\,d\phi\right]}{2H^2}\,,
\ee
while the dilaton field is given by
\be 
\Phi = \frac{\sqrt{3}}{2}\,\ln H\,.
\ee
This procedure yields a consistent four-dimensional solution of the KK theory based on a simple vacuum seed, enhanced by a boost along the compact fifth dimension.

It has been shown that a certain class of transformations can be used to generate the magnetized version of the KK fields \(\left\{ \bar{g}_{\mu\nu}, \bar{A}_\mu, \bar{\Phi} \right\}\), resulting in a new configuration \(\left\{ g_{\mu\nu}, A_\mu, \Phi \right\}\) that includes the effect of an external magnetic field. In \cite{Yazadjiev:2013qna}, a magnetized KK spacetime was constructed that closely parallels the magnetized Kerr--Newman solution in EM theory \cite{Gibbons:2013dna}. That work investigated several physical properties of the solution, including the conserved mass and angular momentum—computed via the Brown--York method—as well as a Smarr-like relation. In the following, we summarize the essential features of the magnetized KK solution that are relevant for our current analysis.

The magnetization procedure begins with a known solution to the equations of motion in KK theory, denoted by \(\left\{ \bar{g}_{\mu\nu}, \bar{A}_\mu, \bar{\Phi} \right\}\), where the seed metric can be expressed as
\be\label{eq.seedmetricKK} 
d\bar{s}^2 = \bar{g}_{\mu\nu} dx^\mu dx^\nu = X_0 \left( d\phi + \omega_0 dt \right)^2 + X_0^{-1} h_{ij} dx^i dx^j\,.
\ee
The accompanying gauge field for the seed solution reads
\be 
{\bar A}_\mu dx^\mu = {\bar A}_t dt + {\bar A}_\phi d\phi \,.
\ee 
From the seed solution above, the corresponding magnetized metric is given by \cite{Yazadjiev:2013qna}
\be\label{metric.magKK}
ds^2 = \frac{X_0}{\Lambda^{1/2}} \left( d\phi + \omega dt \right)^2 + \frac{\Lambda^{1/2}}{X_0} h_{ij} dx^i dx^j\,,
\ee
where the relevant metric functions are defined as
\be 
\omega = \left( 1 + 2b \bar{A}_\phi \right)\omega_0 - 2b \bar{A}_t\,,
\ee
and
\be 
\Lambda = \left( 1 + 2b \bar{A}_\phi \right)^2 + b^2 X_0 e^{2\sqrt{3} \bar{\Phi}}\,.
\ee
The associated gauge field in the magnetized solution is
\be\label{Amu}
A_\mu dx^\mu = \frac{1 + 2b \bar{A}_\phi}{\Lambda} \left( \bar{A}_t dt + \bar{A}_\phi d\phi \right) + \frac{b X_0}{2\Lambda} e^{2\sqrt{3} \bar{\Phi}} \left( \omega_0 dt + d\phi \right)\,,
\ee
and the corresponding dilaton field becomes
\be\label{Dilaton}
\Phi = \frac{\sqrt{3}}{4} \ln \left( \frac{e^{\frac{4}{3} \sqrt{3} \bar{\Phi}}}{\Lambda} \right)\,.
\ee
Here, the parameter \(b\) characterizes the strength of the external magnetic field. This class of magnetized solutions provides a framework to study black hole geometries under the influence of magnetic fields in the context of KK theory.

In principle, any solution set \(\left\{ \bar{g}_{\mu\nu}, \bar{A}_\mu, \bar{\Phi} \right\}\) that satisfies the equations of motion can be magnetized to generate a new solution within the KK framework. In this work, however, we focus specifically on comparing the well-known magnetized Kerr solution in EM theory with its KK counterpart. To facilitate a clear comparison, we adopt the Kerr metric as our seed configuration, rather than a nontrivial KK solution with preexisting gauge and dilaton fields. Since the Kerr spacetime satisfies the vacuum Einstein equations, it trivially fulfills the KK equations of motion (\ref{eq.Rmn})--(\ref{eq.Phi}) under the assumption that the gauge field and dilaton vanish. Equivalently, this choice corresponds to setting the boost parameter \(\alpha = 0\), thereby excluding any intrinsic electric charge. While one could instead begin with a general value of \(\alpha\), resulting in a magnetized, rotating, and electrically charged black hole in KK theory, such an extension introduces additional complexity that is beyond the scope of the present study. Here, our primary goal is to explore the direct analogue of the investigations carried out in \cite{Siahaan:2015xia,Astorino:2015naa}, now within the context of KK theory.

Under this approach, the relevant functions for the seed metric in the form of equation (\ref{eq.seedmetricKK}) are given by
\be 
X_0 = \sin^2\theta \left( \frac{2Mra^2 \sin^2\theta}{r^2 + a^2 \cos^2\theta} + a^2 + r^2 \right),
\ee
\be 
\omega_0 = -\frac{2Mra}{a^4 \cos^2\theta + \left[ r^2(1 + \cos^2\theta) + 2Mr \sin^2\theta \right] a^2 + r^4},
\ee
while the non-vanishing components of \(h_{ij}\) are
\be 
h_{\theta\theta} = a^4 \cos^2\theta + \left[ r^2(1 + \cos^2\theta) + 2Mr \sin^2\theta \right] a^2 + r^4,
\ee
\be 
h_{tt} = -\Delta \sin^2\theta,
\ee
\be 
h_{rr} = \frac{\left( \Delta a^2 \cos^2\theta + r(r^3 + a^2 r + 2Ma^2) \right) \sin^2\theta}{\Delta},
\ee
where \(\Delta = r^2 - 2Mr + a^2\). Substituting the expressions for \(X_0\), \(\omega_0\), and \(h_{ij}\) into the general form (\ref{eq.seedmetricKK}) reproduces the well-known Kerr metric. This sets the stage for investigating its magnetized generalization within the framework of KK theory. Employing the magnetization above to the Kerr metric yields the magnetized Kerr spacetime in the Kaluza-Klein theory, whose metric takes the general form (\ref{metric.magKK}) with the corresponding functions $\Lambda = 1+b^2 X_0$ and $\omega = \omega_0$. 

\begin{figure}[H]
	\centering
	\begin{subfigure}{.5\textwidth}
		\centering
		\includegraphics[width=.8\linewidth]{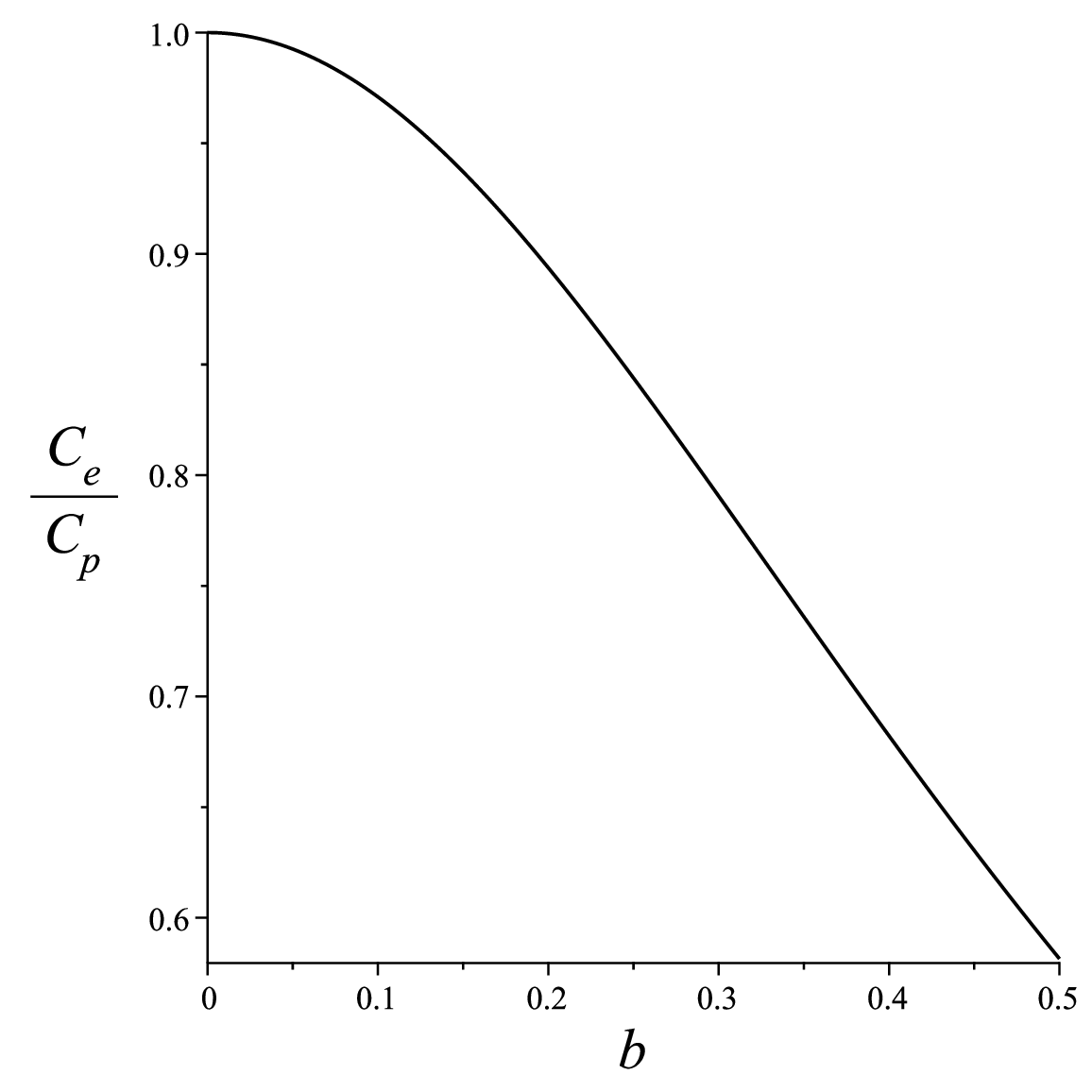}
		\caption{}
		\label{fig:CvsC}
	\end{subfigure}%
	\begin{subfigure}{.5\textwidth}
		\centering
		\includegraphics[width=.8\linewidth]{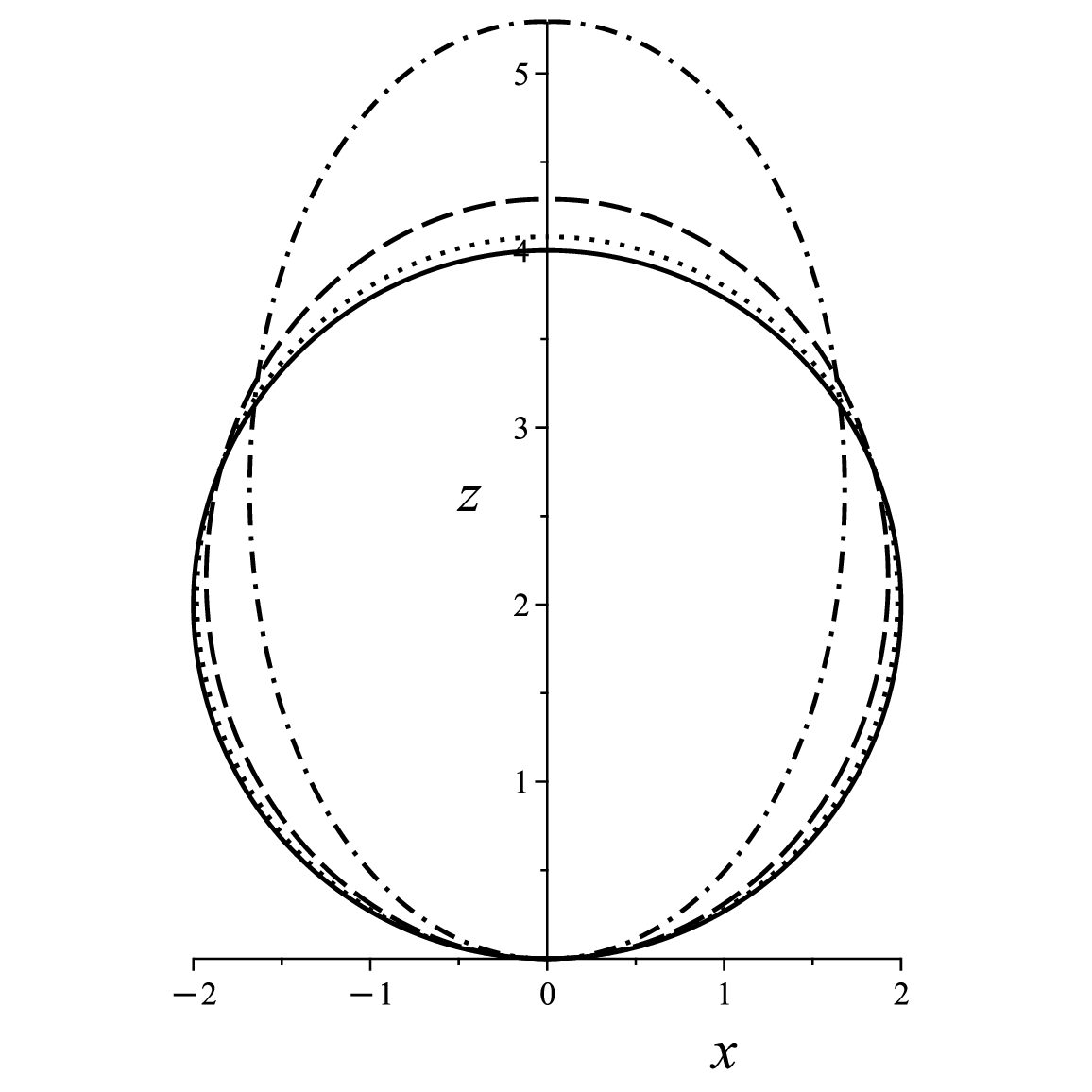}
		\caption{}
		\label{fig:crosssec}
	\end{subfigure}
	\caption{Response of black hole horizons for the variation of $b$. Here we consider the non-rotating case, i.e. magnetized Schwarzschild in KK theory. Fig \ref{fig:CvsC} shows the ratio of polar and equatorial circumferences, as fig. \ref{fig:crosssec} presents the cross-sectional curve of horizon on the plane aligned with the rotational axis of the black hole.}
	\label{fig:Cross}
\end{figure}

A distinctive feature of magnetized black holes is the manner in which their horizons deform in response to an external magnetic field. Within the EM framework, it has been shown that the event horizon becomes increasingly prolate as the strength of the magnetic field increases \cite{Booth:2015nwa}. A similar behavior arises in the context of KK theory, where the deformation can be characterized by comparing the polar and equatorial circumferences of the event horizon \cite{Siahaan:2024fbh}. These quantities are defined as
\be 
C_p = \int\limits_{\theta = 0}^{\pi} \left. \sqrt{g_{\theta\theta}}\, \right|_{r = r_+} d\theta\,,
\ee
and
\be 
C_e = \int\limits_{\phi = 0}^{2\pi} \left. \sqrt{g_{\phi\phi}}\, \right|_{r = r_+,\,\theta = \frac{\pi}{2}} d\phi\,,
\ee
for the polar and equatorial circumferences, respectively. As illustrated in Fig.~\ref{fig:CvsC}, the two circumferences are equal when the magnetic field vanishes (\(b = 0\)), indicating a spherical horizon. However, as the magnetic field strength increases, the polar circumference \(C_p\) grows more rapidly than the equatorial one \(C_e\), signaling that the horizon becomes more prolate under the influence of the external field.

\begin{figure}[H]
	\centering
	\begin{subfigure}{.5\textwidth}
		\centering
		\includegraphics[width=.8\linewidth]{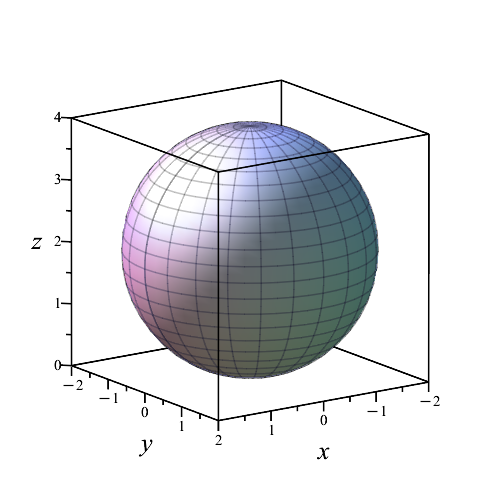}
		\caption{}
		\label{fig:B00}
	\end{subfigure}%
	\begin{subfigure}{.5\textwidth}
		\centering
		\includegraphics[width=.8\linewidth]{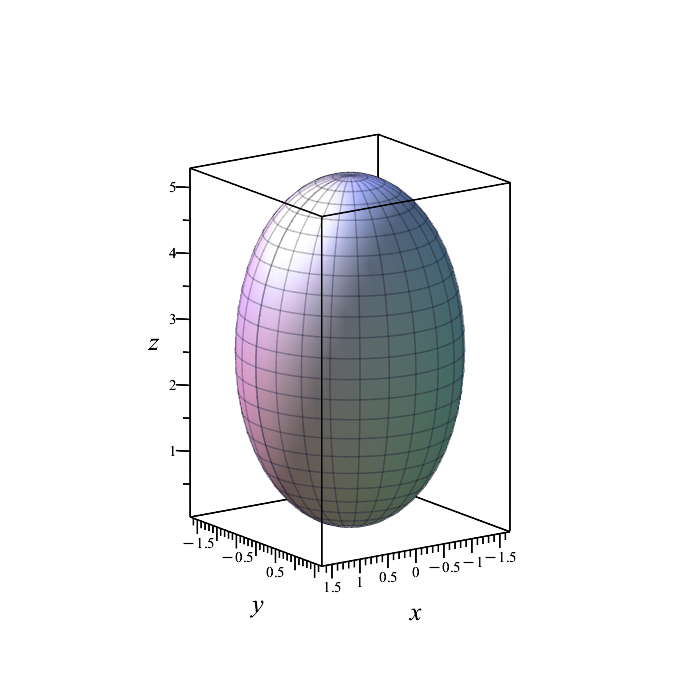}
		\caption{}
		\label{fig:B05}
	\end{subfigure}
	\caption{Illustrations for the horizons embedded in ${\mathbb R}^3$ of the magnetized Schwarzschild black hole in KK theory. Fig. \ref{fig:B00} corresponds to the non-magnetized case, i.e. Schwarzschild black hole, whereas the surface presented in \ref{fig:B05} associates to the $bM=0.5$ case.}
	\label{fig:B0}
\end{figure}

\begin{figure}[H]
	\centering
	\begin{subfigure}{.5\textwidth}
		\centering
		\includegraphics[width=.8\linewidth]{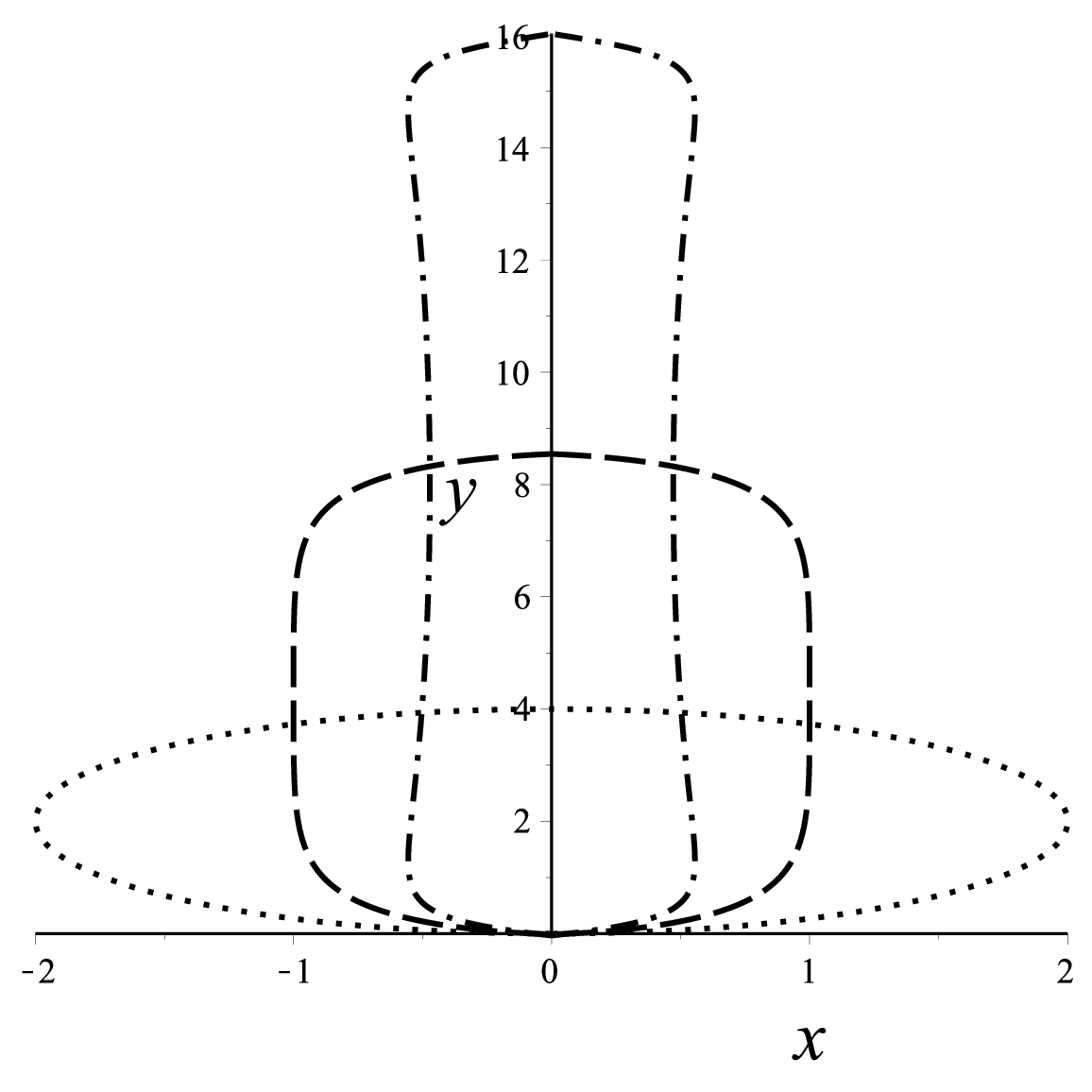}
		\caption{}
		\label{fig:crossecEM}
	\end{subfigure}%
	\begin{subfigure}{.5\textwidth}
		\centering
		\includegraphics[width=.6\linewidth]{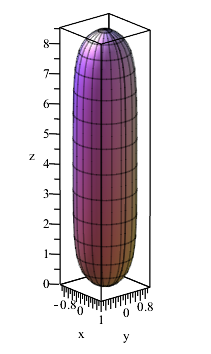}
		\caption{}
		\label{fig:B05EM}
	\end{subfigure}
	\caption{The horizons of a magnetized Schwarzschild black hole in Einstein-Maxwell theory, embedded in $\mathbb{R}^3$, are illustrated in this figure. Figure \ref{fig:crossecEM} provides a cross-sectional view of the horizon in the plane aligned with the black hole's rotational axis. Different magnetic field strengths are represented by distinct line styles: dots ($bM=0$), dashed lines ($bM=0.5$), and dashed-dotted lines ($bM=0.9$). An unconstrained scaling is used in Figure \ref{fig:crossecEM} to enhance the visibility of the negative curvature near the equator for the $bM=0.9$ case, which distorts the initially circular horizon (dots) into an ellipsoidal shape. Figure \ref{fig:B05EM} offers a dedicated illustration of the horizon when the magnetic field strength is $bM=0.5$.}
	\label{fig:BEM}
\end{figure}

To provide a clearer visualization, we adopt the approach of \cite{Booth:2015nwa} to illustrate how the black hole horizon deforms as the external magnetic field strength increases. For simplicity, we restrict the analysis to the non-rotating case, as applied in Figures~\ref{fig:Cross} -- \ref{fig:BEM}. As shown in Figures~\ref{fig:Cross} and~\ref{fig:B0}, the horizon of a magnetized black hole in KK theory becomes increasingly prolate under stronger magnetic fields, exhibiting a similar qualitative deformation as observed in the EM case, illustrated in Figure~\ref{fig:BEM}. However, one key distinction emerges: while both EM and KK black holes show horizon elongation along the polar axis, only in the EM case do we observe a transition in the equatorial curvature from positive to negative as the magnetic field grows. This feature, absent in the KK counterpart, highlights a fundamental difference between the two frameworks. Detailed analyses of the Ricci scalar for the horizon surface in both EM and KK theories can be found in \cite{Siahaan:2024fbh}; here, we focus on illustrative comparisons to emphasize that not all properties of magnetized black holes in EM theory carry over to the KK setting.

\section{Black hole thermodynamics}\label{sec.thermo}

Since one of the central topics in the next section involves the entropy of magnetized black holes, we begin here with a brief discussion of the thermodynamic properties of magnetized Kerr black holes in KK theory. This section closely follows the analysis presented in \cite{Yazadjiev:2013qna}. It is important to note that while our focus is on the magnetized Kerr solution within the KK framework, whereas the work in \cite{Yazadjiev:2013qna} investigates a broader class of rotating and charged KK black holes, analogous to the Kerr--Newman solution in EM theory.

In the construction of \cite{Yazadjiev:2013qna}, the presence of a boost parameter \(\alpha\) gives rise to an electric charge \( Q \), with the relation
\be
Q = M \sinh\alpha \cosh\alpha\,.
\ee
However, in our case, where the seed solution is the neutral Kerr black hole, the boost parameter is fixed such that \(\cosh\alpha = 1\), leading to \(Q = 0\). As a result, the solution under consideration remains electrically neutral. Employing the standard Komar integral formalism, the conserved quantities associated with the Kerr spacetime are identified as the ADM mass \(M\) and the angular momentum \(J = Ma\), which will play a role in the thermodynamic analysis that follows.

To verify the location of the event horizon, we examine the \((r,r)\) component of the covariant metric tensor, which is given by
\be
g_{rr} = \frac{1}{\Delta} \sqrt{ \left\{ \Sigma + b^2 \sin^2\theta \left[ \Sigma \left(a^2 + r^2\right) + 2a^2 M r \sin^2\theta \right] \right\} \Sigma },
\ee
where \(\Sigma = r^2 + a^2 \cos^2\theta\). From the explicit form of \(g_{rr}\), we observe that the magnetic field parameter \(b\) carries inverse mass dimension, consistent with the analogous parameter in the EM theory of magnetized black holes \cite{Siahaan:2021ags}. The expression under the square root is manifestly positive and non-vanishing, implying that the presence of the external magnetic field does not affect the location of the event horizon. Therefore, the horizon structure of the magnetized Kerr black hole in KK theory remains unchanged, and the radial positions of the inner and outer horizons are determined by the roots of \(\Delta = r^2 - 2Mr + a^2\), yielding
\be
r_\pm = M \pm \sqrt{M^2 - a^2}\,,
\ee
which are identical to those of the unmagnetized Kerr solution.

The area of the outer event horizon can be computed via the standard formula
\be
A_H = \int\limits_{\phi = 0}^{2\pi} d\phi \int\limits_{\theta = 0}^{\pi} \left. \sqrt{g_{\phi\phi} \, g_{\theta\theta}} \right|_{r = r_+} d\theta,
\ee
where the relevant metric components are given by
\be
g_{\theta\theta} = \sqrt{ \left\{ \Sigma + b^2 \sin^2\theta \left[ \Sigma \left(a^2 + r^2\right) + 2a^2 M r \sin^2\theta \right] \right\} \Sigma },
\ee
and
\be
g_{\phi\phi} = \frac{ \sin^2\theta \left[ \Sigma \left(a^2 + r^2\right) + 2a^2 M r \sin^2\theta \right] }{ \sqrt{ \left\{ \Sigma + b^2 \sin^2\theta \left[ \Sigma \left(a^2 + r^2\right) + 2a^2 M r \sin^2\theta \right] \right\} \Sigma } }.
\ee
Evaluating the integral yields
\be \label{eq.Area}
A_H = 8\pi M r_+,
\ee
which is precisely the area of the horizon of a non-magnetized Kerr black hole. This invariance of the horizon area under magnetization is a noteworthy feature, and it is also observed in the EM framework. For instance, in the case of the magnetized Kerr--Newman--Taub--NUT black hole, the horizon area remains identical to that of its unmagnetized counterpart \cite{Ghezelbash:2021xvc,Ghezelbash:2021lcf}.

Furthermore, the area expression in Eq.~\eqref{eq.Area} is independent of the external magnetic field parameter \(b\). This invariance of the horizon area under magnetization is also observed in the EM theory. In other words, although the geometry of the black hole horizon may be deformed by varying \(b\), such deformation does not lead to a change in the total area of the horizon.

Recall that the seed solution prior to magnetization is the electrically neutral Kerr spacetime, and the resulting configuration is the magnetized Kerr black hole in KK theory. The physical electric charge of the magnetized solution can be computed via the integral
\be 
Q = \frac{1}{4\pi} \int_{S} e^{-2\sqrt{3}\Phi} \star \mathbf{F} = -2bMa\,,
\ee
where \(\star\) denotes the Hodge dual operator, \(\mathbf{F}\) is the two-form field strength tensor, and the integration is performed over a two-dimensional surface \(S\) of constant time and radius. Notably, the magnitude of this induced electric charge coincides with that of the magnetized Kerr black hole in EM theory \cite{Astorino:2015naa}. Note that a static black hole does not acquire electric charge merely due to the presence of an external magnetic field. The magnetic field considered here is purely external and is not sourced by any intrinsic magnetic or dyonic charge of the black hole. In this work, we focus exclusively on electrically neutral black holes without magnetic monopole contributions.

The computation of mass and angular momentum for the magnetized Kerr black hole in KK theory cannot be carried out using the standard Komar integral method, as is typically done in the non-magnetized case. This limitation arises due to the non-asymptotically flat nature of the magnetized Kerr geometry, a feature common to both EM and KK frameworks. In \cite{Yazadjiev:2013qna}, the author employed the quasilocal formalism to evaluate the conserved quantities. It was found that the quasilocal mass and angular momentum of the magnetized solution exactly coincide with those of the corresponding non-magnetized black hole.

Given that the horizon area also remains unchanged under magnetization and that we consider a seed solution with vanishing electric charge, the Smarr-like relation for the magnetized Kerr black hole in KK theory takes the familiar form \cite{Yazadjiev:2013qna}
\be \label{eq.MassSmarr}
M = \frac{\kappa}{4\pi} A_+ + 2\Omega_+ J\,,
\ee
where the surface gravity is given by
\be \label{eq.kappa}
\kappa = \frac{r_+ - r_-}{4 M r_+}\,,
\ee
and the angular velocity at the horizon reads
\be \label{eq.OmH}
\Omega_H = \frac{a}{2 M r_+}\,.
\ee

As expected, Eq.~\eqref{eq.MassSmarr} is identical to the Smarr relation for the unmagnetized Kerr solution in vacuum Einstein gravity. Consequently, the first law of black hole thermodynamics for the magnetized Kerr solution in KK theory retains its standard form:
\begin{equation} \label{eq.dM}
\delta M = \frac{\kappa}{2\pi} \delta S_{BH} + \Omega_H \delta J\,,
\end{equation}
where the Bekenstein--Hawking entropy follows the usual quarter-area law, 
\be \label{eq.Sbh}
S_{BH} = \frac{A_{H}}{4}\,,
\ee 
where horizon's area $A_H$ is given in eq. (\ref{eq.Area}). From this relation, the Hawking temperature associated with the magnetized Kerr black hole in KK theory is given by
\begin{equation} \label{eq.temp}
T_H = \frac{\kappa}{2\pi},
\end{equation}
which coincides exactly with the Hawking temperature of the neutral Kerr black hole. This invariance of temperature under magnetization is also observed in the EM theory \cite{Siahaan:2015xia,Astorino:2015naa,Siahaan:2021ags,Siahaan:2021uqo,Ghezelbash:2021xvc}, further emphasizing the robustness of thermodynamic properties in the presence of external magnetic fields.

To conclude this section, we highlight another key distinction between magnetized black holes in KK theory and their EM counterparts. In the EM case, the black hole entropy computed via the standard quarter area law depends explicitly on the strength of the external magnetic field \cite{Siahaan:2015xia,Astorino:2015naa}. In contrast, the entropy of magnetized black holes in KK theory is entirely independent of the magnetic field parameter \cite{Yazadjiev:2013qna}; it exactly matches that of the unmagnetized configuration. This property will play a significant role in the Kerr/CFT analysis discussed in the following sections.

\section{Near horizon conformal symmetry and Cardy formula}\label{sec.nearhor}

To uncover the structure of the extremal black hole's near-horizon region, one typically performs a suitable coordinate transformation that zooms in on the immediate vicinity of the horizon. To do so, we transform the \(r\), \(t\), and \(\phi\) coordinates by ensuring that the leading-order terms in the metric and fields manifest with enhanced symmetry. Below, we show in detail how this procedure applies to both the metric and the accompanying non--gravitational fields into a near-horizon form, revealing the underlying conformal symmetry of the extremal solution.

The near horizon geometry is obtained after performing the coordinate transformation
\begin{equation} \label{eq.nhek}
r \to a + \varepsilon {\sqrt{2}a} y~~,~~t \to {\sqrt{2}a} \frac{\tau }{\varepsilon }~~,~~\phi \to \varphi + \frac{\sqrt{2}a \tau}{2m \varepsilon} \,,
\end{equation}
which transform the metric, gauge field, and dilaton field above to
\begin{equation} \label{eq.nearmetric}
ds^2 = \Gamma (\theta) \left( { - y^2 d\tau ^2 + \frac{{dy^2 }}{{y^2 }} + d\theta^2} \right) + \beta \left(\theta\right)\left( {d\varphi + {y}d\tau } \right)^2 \,,
\end{equation}
\begin{equation}
A_\mu dx^\mu = - \frac{{2a^6 b\left( {1 + \cos ^2 \theta } \right)\sin ^2 \theta }}{{\Gamma \left( \theta \right)^2 }}\left( {yd\tau + d\varphi } \right)\,,
\end{equation}
\begin{equation}
\Phi = - \frac{{\sqrt 3 }}{4}\ln \left( {\frac{{\Gamma \left( \theta \right)^2 }}{{1 + \cos ^2 \theta }}} \right)\,,
\end{equation}
where
\begin{equation}
\Gamma \left( \theta \right) = a^2 \sqrt {\left( {1 + \cos ^2 \theta + 4a^2 b^2 \sin ^2 \theta } \right)\left( {1 + \cos ^2 \theta } \right)}\,,
\end{equation}
and
\begin{equation}
\beta \left( \theta \right) = \frac{{4a^4 \sin ^2 \theta }}{{\Gamma \left( \theta \right)}}\,,
\end{equation}
after considering the extremal limit $M=a$ and taking $\varepsilon\to 0$.
These fields obey the equations of motion in eqs. (\ref{eq.Rmn}) -- (\ref{eq.Phi}). Note that the resulting near-horizon of extremal black hole geometries takes the form that has been discussed repeatedly in the context of Kerr/CFT holography \cite{Compere:2012jk}. The near-horizon metric (\ref{eq.nearmetric}) takes the typical AdS$_2\times$dS$^2$ warped and twisted product, and possesses the $SL(2,{\mathbb R}) \times U(1)$ isometry.

The procedure for computing the central charge associated with the conformal symmetry of extremal KK black holes was established in \cite{Azeyanagi:2008kb,Li:2010ch}, and it can be applied here in a similar fashion. This approach involves imposing appropriate boundary conditions on the diffeomorphisms of the fields, such that the variations satisfy ${\cal L}_{\zeta_n} g_{\mu\nu} = h_{\mu\nu}$, ${\cal L}_{\zeta_n} A_{\mu} = \delta A_{\mu}$, and ${\cal L}_{\zeta_n} \Phi = \delta \Phi$, where the transformations are generated by the vector field $\zeta_n$.
\begin{equation}
{\zeta_n} = -e^{-in\varphi} \frac{\partial}{\partial \varphi}- iny e^{-in\varphi} \frac{\partial}{\partial y}\,,
\end{equation}
which satisfies the Virasoro algebra
\begin{equation}
i \left[\zeta_m,\zeta_n\right] = \left(m-n\right) \zeta_{n+m}\,.
\end{equation}
Furthermore, each diffeomorphism is related to a conserved charge $Q_\zeta$ by adopting the well-known Barnich-Brandt method \cite{Barnich:2001jy,Compere:2012jk}. Between these charges, one can establish the Dirac brackets
\begin{equation}
\left\{ {Q_\zeta ,Q_\xi } \right\} = Q_{\left[ {\zeta ,\xi } \right]} - \int\limits_{\partial M} {{\bf K}_\zeta }
\end{equation}
where the last term is known as the central term. Explicitly, the integrand of the central term can be expressed as
\begin{equation}
{\bf K}_\zeta \left[ {h,g} \right] = \frac{1}{{64\pi }}\varepsilon _{\alpha \beta \mu \nu } K_\zeta ^{\alpha \beta } dx^\mu \wedge dx^\nu \,,
\end{equation}
where
\begin{equation}
K_\zeta ^{\alpha \beta } = \zeta ^\beta \nabla ^\alpha h - \zeta ^\beta \nabla _\sigma h^{\alpha \sigma } + \frac{h}{2}\nabla ^\beta \zeta ^\alpha - h^{\beta \sigma } \nabla _\sigma \zeta ^\alpha + \zeta _\sigma \nabla ^\beta h^{\alpha \sigma } - \left( {\alpha \to \beta } \right) \,,
\end{equation}
and $\epsilon_{0123} = \sqrt{-{\det} \left[{ g}_{\alpha\beta}\right]}$. In the calculations above, the covariant derivative together with the lowering/raising index operations are done by using the metric tensor for the near-horizon of extremal geometry (\ref{eq.nearmetric}). Explicitly, the calculation of the central term can be written as
\begin{equation}
\int\limits_{\partial M} {K_{\zeta _m } } \left[ {L_{\zeta _n } g,g} \right] \sim - \frac{icm^3}{{12}}\delta _{m, - n} \,.
\end{equation}
A general formula to compute the central charge which consists of the near-horizon of extremal metric functions as appeared in (\ref{eq.nearmetric}) has been given in \cite{Compere:2012jk}, namely
\begin{equation} \label{eq.centralGen}
c = 3\int\limits_{ 0}^\pi {d\theta\sqrt {\Gamma \left( \theta \right) \beta \left( \theta \right)} } \,.
\end{equation}
For the near-horizon of extremal magnetized Kerr black hole in KK theory as indicated in eq. (\ref{eq.nearmetric}), the central charge is found to be
\begin{equation} \label{eq.central}
c = 12a^2\,.
\end{equation}
This central charge coincides with that of the extremal Kerr black hole, as the external magnetic field introduces no contribution to its value. Moreover, unlike in the EM case \cite{Siahaan:2015xia,Astorino:2015naa}, the central charge here remains manifestly positive, ensuring that the dual two-dimensional CFT is unitary.

To complete the Kerr/CFT correspondence prescription for reproducing the extremal black hole entropy via the Cardy formula, one must determine the generalized temperature associated with the Frolov--Thorne vacuum. This is achieved by considering the eigenmodes of a scalar field characterized by frequency \(\omega\) and azimuthal quantum number \(m\):
\begin{equation}
\exp\left(-i\omega t + i m \phi\right) = \exp\left[-i\left(\omega - \frac{m}{2M}\right)\frac{\sqrt{2}a \tau}{\varepsilon} + i m \varphi\right],
\end{equation}
where the coordinate transformation from Eq.~\eqref{eq.nhek} has been applied. Recognizing the right-hand side as
\(\exp(-i n_R \tau + i n_L \varphi)\), as in \cite{Guica:2008mu}, we identify the left- and right-moving quantum numbers as
\begin{equation}
n_R = \left(\omega - \frac{m}{2a}\right)\frac{\sqrt{2}a}{\varepsilon}, \qquad n_L = m.
\end{equation}

The Boltzmann factor for the scalar field modes can then be written in the form \cite{Compere:2012jk}
\begin{equation}
\exp\left(\frac{\omega - \Omega_H m}{T_H}\right) = \exp\left(-\frac{n_L}{T_L} - \frac{n_R}{T_R}\right),
\end{equation}
from which the left and right temperatures are extracted:
\begin{equation}
T_R = -\frac{\sqrt{2}a T_H}{\varepsilon},
\end{equation}
\begin{equation}
T_L = \left[\lim_{r_h \to a} \frac{T_H}{\frac{1}{2a} - \Omega_H}\right] = -\left.\frac{\partial T_H / \partial r_h}{\partial \Omega_H / \partial r_h}\right|_{r_h = a} = \frac{1}{2\pi},
\end{equation}
where \(\Omega_H\) denotes the angular velocity at the outer horizon, given in Eq.~\eqref{eq.OmH}, and \(T_H\) is the Hawking temperature evaluated at \(r_h = r_+\), as given in Eq.~\eqref{eq.temp}. In the extremal limit \(a = M\), the Hawking temperature vanishes, \(T_H = 0\), implying that the right-moving temperature also vanishes, \(T_R = 0\), while the left-moving temperature remains finite:
\begin{equation} \label{eq.TL}
T_L = \frac{1}{2\pi}.
\end{equation}
It is important to note that the extremality condition \(T_H \to 0\) and the near-horizon limit \(\varepsilon \to 0\) are logically distinct; the vanishing of \(T_R\) does not require \(\varepsilon \to 0\), as discussed in \cite{Guica:2008mu}. 

Finally, applying the Cardy formula
\begin{equation}
S_{\text{CFT}} = \frac{\pi^2}{3} c T_L,
\end{equation}
with the central charge \(c\) given in Eq.~\eqref{eq.central} and the left-moving temperature \(T_L\) in Eq.~\eqref{eq.TL}, yields the microscopic entropy of the extremal magnetized Kerr black hole:
\begin{equation}
S_{\text{ext.}} = 2\pi a^2,
\end{equation}
which exactly matches one-quarter of the horizon area in the extremal limit, as given by Eq.~\eqref{eq.Area}. This agreement demonstrates that the Kerr/CFT correspondence successfully reproduces the Bekenstein--Hawking entropy for the extremal magnetized Kerr black hole in KK theory, thereby validating its applicability in this extended framework.

\section{Hidden conformal symmetry and microscopic entropy}\label{sec.hiddenconf}

To complement the extremal Kerr/CFT correspondence discussed in the previous section, we now explore the hidden conformal symmetry of the black hole. This approach enables a microscopic derivation of the black hole entropy even in the non-extremal regime. In their seminal work \cite{Castro:2010fd}, Castro et al.\ demonstrated that a hidden two-dimensional conformal symmetry arises in the near-region dynamics of the scalar field propagating in a Kerr black hole background. Remarkably, this conformal symmetry is not manifest in the spacetime geometry itself but instead emerges from the structure of the low-frequency, near-region solution to the massless scalar wave equation. Furthermore, the authors showed that the scalar-Kerr scattering amplitudes can be matched with the correlation functions of a 2D conformal field theory (CFT), and that the Cardy formula reproduces the Bekenstein--Hawking entropy. These results significantly bolster the Kerr/CFT conjecture, extending its applicability beyond the extremal limit.

In the series of works on the Kerr/CFT correspondence for magnetized black holes within EM theory \cite{Siahaan:2015xia,Astorino:2015naa,Siahaan:2021ags,Ghezelbash:2021lcf,Ghezelbash:2021xvc}, investigations have primarily focused on the extremal limit, where the microscopic entropy derived from the Cardy formula matches the macroscopic Bekenstein--Hawking entropy. A key reason for this restriction lies in the non-separability of the KG equation even in the massless, neutral case in the background of magnetized black holes in EM theory. This non-separability obstructs the identification of a hidden conformal structure in the non-extremal regime. However, as we demonstrate below, the situation is notably different for the magnetized black hole in KK theory. In this case, we find that the KG equation for a massless neutral scalar field remains separable into radial and angular components, thus opening the possibility to uncover hidden conformal symmetry beyond extremality.

As a starting point, let us consider the KG equation for a neutral, massless scalar field perturbation:
\[
\nabla_\mu \nabla^\mu \Phi = 0,
\]
and adopt the standard mode decomposition ansatz \(\Phi(t, r, \theta, \phi) = e^{-i\omega t + i m \phi} R(r) S(\theta)\). This separation is permitted due to the stationary and axisymmetric Killing symmetries of the magnetized KK spacetime considered in this work. After straightforward algebraic manipulations, the KG equation reduces to two decoupled ordinary differential equations. The radial equation takes the form
\[
\frac{d}{dr} \left( \Delta \frac{dR(r)}{dr} \right) + \left[ \frac{(2M r \omega - a m)^2}{\Delta} + \omega^2 \Delta + 4M\omega^2 r - b^2 r^2 m^2 - \lambda \right] R(r) = 0,
\]
while the angular equation becomes
\[
\frac{1}{\sin\theta} \frac{d}{d\theta} \left( \sin\theta \frac{dS(\theta)}{d\theta} \right) - \left[ \omega^2 a^2 \sin^2\theta + \frac{m^2}{\sin^2\theta} \left(1 + a^2 b^2 \cos^2\theta \sin^2\theta \right) - \lambda \right] S(\theta) = 0.
\]
It is important to emphasize that these equations are obtained without imposing any restriction on the strength of the external magnetic field parameter \(b\). This stands in contrast to the case of magnetized Kerr black holes in EM theory, where separability of the KG equation— even for neutral, massless scalars— typically requires assuming the weak-field limit \(bM \ll 1\) \cite{Konoplya:2007yy}.

In the equations above, \(\lambda\) is a separation constant. Following the prescription in \cite{Guica:2008mu}, the radial equation can be recast in the form
\begin{equation} \label{eq.Rad}
\frac{d}{dr} \left( \Delta \frac{dR(r)}{dr} \right) + \left[ \frac{(2M r_+ \omega - a m)^2}{(r - r_+)(r_+ - r_-)} - \frac{(2M r_- \omega - a m)^2}{(r - r_-)(r_+ - r_-)} + f(r) \right] R(r) = \lambda R(r),
\end{equation}
where the residual function is given by
\[
f(r) = \omega^2 \Delta + 4M(r + M) \omega^2 - b^2 r^2 m^2.
\]

To reveal the hidden conformal symmetry probed by a test scalar field in the Kerr/CFT framework, we adopt the standard conditions used in previous literature. Specifically, we consider the low-frequency limit \(M\omega \ll 1\) and the near-region approximation \(r\omega \ll 1\), as originally introduced in \cite{Castro:2010fd}. In the context of magnetized spacetimes, it is also important to account for the contribution of the external magnetic field to the geometry. A physically relevant assumption, widely supported in the literature, is that the astrophysical magnetic fields are weak, satisfying \(bM \ll 1\). Interestingly, this weak-field condition is fully consistent with our objective of isolating the hidden conformal symmetry in the near-horizon dynamics of the test scalar. Under the combined assumptions of low frequency, near-region limit, and weak magnetic field, the function \(f(r)\) becomes negligible in Eq.~\eqref{eq.Rad}. As a result, the radial equation reduces to a form that exhibits an \(SL(2, \mathbb{R})\) symmetry structure, allowing for the identification of hidden conformal symmetry in the background of the magnetized Kerr black hole in KK theory.

The subsequent analysis closely parallels the standard treatment of hidden conformal symmetry and will not be repeated in full detail here. For comprehensive derivations, we refer the reader to foundational works such as \cite{Castro:2010fd,Guica:2008mu}. In the absence of the \( f(r) \) contribution in the radial equation \eqref{eq.Rad}, the equation can be identified as an eigenvalue equation for the quadratic Casimir operator of \( SL(2,\mathbb{R}) \). More precisely, there exist two independent sets of \( SL(2,\mathbb{R}) \) generators acting on the solution space of the scalar field, defined as follows \cite{Guica:2008mu}:
\begin{align}
H_{+} &= i\,e^{-2\pi T_R \phi} \left( \sqrt{\Delta}\,\partial_r + \frac{r - M}{2\pi T_R \sqrt{\Delta}}\,\partial_\phi + \frac{2 T_L (M r - a^2)}{T_R \sqrt{\Delta}}\,\partial_t \right), \notag \\
H_{0} &= \frac{i}{2\pi T_R}\,\partial_\phi + \frac{2i M T_L}{T_R}\,\partial_t, \notag \\
H_{-} &= i\,e^{2\pi T_R \phi} \left( -\sqrt{\Delta}\,\partial_r + \frac{r - M}{2\pi T_R \sqrt{\Delta}}\,\partial_\phi + \frac{2 T_L (M r - a^2)}{T_R \sqrt{\Delta}}\,\partial_t \right),
\end{align}
and
\begin{align}
\bar{H}_{+} &= i\,e^{-2\pi T_L \phi + t/2M} \left( \sqrt{\Delta}\,\partial_r - \frac{a}{\sqrt{\Delta}}\,\partial_\phi + \frac{2 M r}{\sqrt{\Delta}}\,\partial_t \right), \notag \\
\bar{H}_{0} &= -2iM\,\partial_t, \notag \\
\bar{H}_{-} &= i\,e^{2\pi T_L \phi - t/2M} \left( -\sqrt{\Delta}\,\partial_r - \frac{a}{\sqrt{\Delta}}\,\partial_\phi - \frac{2 M r}{\sqrt{\Delta}}\,\partial_t \right).
\end{align}

These generators satisfy the \( SL(2,\mathbb{R}) \) Lie algebra:
\begin{equation}
[H_0, H_{\pm}] = \mp i H_{\pm}, \quad [H_{-}, H_{+}] = -2i H_0,
\end{equation}
\begin{equation}
[\bar{H}_0, \bar{H}_{\pm}] = \mp i \bar{H}_{\pm}, \quad [\bar{H}_{-}, \bar{H}_{+}] = -2i \bar{H}_0.
\end{equation}
Recognizing the separation constant as \(\lambda = l(l+1)\), the conformal weights of the scalar field \(\Phi\) are identified as \( h_L = h_R = l \), consistent with the representation theory of \( SL(2,\mathbb{R})_L \times SL(2,\mathbb{R})_R \). The corresponding left and right temperatures are given by
\begin{equation}
T_L = \frac{r_+ + r_-}{4\pi a}, \quad T_R = \frac{r_+ - r_-}{4\pi a}.
\end{equation}

Having established the presence of hidden conformal symmetry associated with the non-extremal magnetized KK black hole, we now turn to the corresponding microscopic entropy calculation. As argued in \cite{Guica:2008mu}, it is reasonable to assume that the central charge remains unchanged from its extremal value, namely \( c_L = c_R = 12a^2 \). Under this assumption, and invoking the validity of the Cardy formula beyond extremality, the microscopic entropy can be computed as
\begin{equation}
S_{\rm{non\text{-}ext.}} = \frac{\pi^2}{3} \left( c_L T_L + c_R T_R \right) = 2\pi M r_+\,.
\end{equation}
This result precisely reproduces the Bekenstein--Hawking entropy of the non-extremal magnetized KK black hole in eq. (\ref{eq.Sbh}), which, as previously noted, coincides with that of the unmagnetized Kerr black hole.

In the broader context of hidden conformal symmetry and the Kerr/CFT correspondence, the dual description also encompasses scalar field scattering in the black hole background. While the detailed analysis parallels that of \cite{Guica:2008mu}, and thus will not be repeated here, it is worth noting that the outcome confirms the duality. Specifically, the absorption cross section for a low-frequency, neutral scalar field in the near region of a weakly magnetized KK black hole can be precisely reproduced using the standard two-dimensional CFT expression, further supporting the holographic interpretation. 

\section{Conclusion}

In this work, we have investigated several novel aspects of magnetized black holes within the framework of KK theory. A central result is that the near-horizon geometry of the extremal black hole exhibits a warped and twisted AdS$_2 \times$ dS$_2$ structure, reaffirming the applicability of the Kerr/CFT correspondence in this context. We have demonstrated that the entropies of both extremal and non-extremal magnetized Kerr black holes in KK theory can be consistently reproduced through Kerr/CFT holography: the extremal case via the emergence of near-horizon conformal symmetry, and the non-extremal case through hidden conformal symmetry revealed by the separability of the KG equation. Notably, we find that the central charge remains strictly positive and independent of the external magnetic field, ensuring that the dual two-dimensional CFT describing the extremal black hole is unitary -- a feature that stands in contrast to the corresponding scenario in EM theory.

A particularly surprising result is that a massless neutral scalar field in the non-extremal magnetized Kerr--KK background obeys a fully separable KG equation---unlike in the EM counterpart \cite{Konoplya:2007yy}, where separability fails and hidden conformal symmetry only emerges near extremality \cite{Siahaan:2024dso}. Notably, in the KK setup, separability holds generally and does not depend on the assumption of a weak magnetic field ($bM \ll 1$). This broader applicability affirms the physical significance of the hidden conformal symmetry in the low-frequency near-region limit.

These findings open several promising directions for future research. First, a natural extension is to explore the Kerr/CFT correspondence within a broader class of KK black hole solutions. This could include the incorporation of additional parameters such as the boost \(\alpha\), NUT charge, or acceleration. Second, it would be valuable to examine whether the separability of the KG equation observed here extends to more general perturbations, such as those involving massive or charged scalar fields. A positive outcome would indicate an even richer underlying symmetry and could enable detailed analyses of quasinormal modes, similar to those carried out in the EM framework \cite{Konoplya:2007yy}. Finally, the issue of superradiant instability in this KK context presents an interesting aspect for investigation, especially given the existing literature on its counterpart in the EM theory \cite{Konoplya:2008hj,Brito:2014nja}.

\section*{Acknowledgement}

This work was supported by LPPM-UNPAR through the Penelitian Publikasi Internasional Bereputasi funding scheme.

\end{document}